\documentclass{aa}
\usepackage{graphicx}
\usepackage{txfonts}
\usepackage{natbib}

\begin{document} 

\title{Study of the photon-induced formation and subsequent desorption of CH$_{3}$OH and H$_{2}$CO in interstellar ice analogs}

\author{R. Mart\'in-Dom\'enech \inst{\ref{inst1}}  \and G.M. Mu\~noz Caro \inst{\ref{inst1}} \and G.A. Cruz-D\'iaz \inst{\ref{inst2},\ref{inst3}}
}

\institute{Centro de Astrobiolog\'ia (INTA-CSIC), Ctra. de Ajalvir, km 4, Torrej\'on de Ardoz, 28850 Madrid, Spain
\email{rmartin@cab.inta-csic.es}\label{inst1}
\and 
NASA Ames Research Center, Moffett Field, Mountain View, CA 94035, USA\label{inst2}
\and
Bay Area Environmental Research Institute, Petaluma, CA 94952, USA\label{inst3}
}

\date{}

\abstract
{Methanol and formaldehyde are two simple organic molecules that are ubiquitously detected in the interstellar medium. 
An origin in the solid phase and a subsequent nonthermal desorption into the gas phase is often invoked to explain their abundances 
in some of the environments where they are found. 
Experimental simulations under astrophysically relevant conditions have been carried out in the past four decades in order to find a suitable 
mechanism for that process.
}
{
We 
explore the \textit{in situ} formation and subsequent 
photon-induced desorption of these species, 
studying 
the UV photoprocessing of pure ethanol ice, and a more realistic binary H$_{2}$O:CH$_{4}$ ice analog.} 
{Experimental simulations were performed in an ultra-high vacuum chamber. 
Pure ethanol and binary H$_{2}$O:CH$_{4}$ 
ice samples deposited onto an infrared transparent window at 8 K were UV-irradiated using a microwave-discharged hydrogen flow lamp. 
Evidence of photochemical production of these two species and subsequent UV-photon-induced desorption into the gas phase were searched for 
by means of a Fourier transform infrared spectrometer and a quadrupole mass spectrometer, respectively.}
{Formation of CH$_{3}$OH was only observed during photoprocessing of the H$_{2}$O:CH$_{4}$ ice analog, 
but no photon-induced desorption was detected. 
Photochemical production of H$_{2}$CO was observed in both series of experiments. 
Photochemidesorption of formaldehyde, 
i.e., photon-induced formation on the ice surface and inmediate desorption, 
was observed, 
with a yield of $\sim$6 x 10$^{-5}$ (molecules/incident photon) in the case of the pure ethanol 
ice experiments, and $\sim$4.4 x 10$^{-5}$ (molecules/incident photon) when the H$_{2}$O:CH$_{4}$ ice analogs were 
photoprocessed. 
Photoprocessing of the ice analogs lead to formation of other species.  
Some of them were also found to desorb upon UV irradiation.}  
{While certain C-bearing species, in particular H$_{2}$CO, were found to desorb upon irradiation, nonthermal desorption of CH$_{3}$OH was not 
observed.  
So far, there is no experimental evidence of any efficient CH$_{3}$OH desorption induced by UV photons. 
On the other hand, the observed photon-induced desorption of H$_{2}$CO could account for the total formaldehyde abundance observed in the 
Horsehead PDR.}


\titlerunning{UV formation and desorption of simple organic molecules}

\maketitle

\section{Introduction}
\label{intro}
More than 180 molecules have been detected to date in the interstellar medium (ISM). 
Simple organic molecules 
(hydrogenated species 
with just one carbon atom 
and another heavy element)
are proposed as precursors of more complex organic species with prebiotic interest. 
Two of the most studied simple organic molecules
 are CH$_{3}$OH and H$_{2}$CO. 

These species 
are found in a variety of interstellar environments, such as molecular clouds and hot cores, and in comets inside our solar system, 
with abundances relative to H$_{2}$ ranging from 10$^{-6}$ to 10$^{-9}$
\citep[see, e.g.,][]{sutton95,wootten96,ceccarelli00,ehrenfreund02,maret04,smith04,young04,araya07,leurini10,bergman11,guzman11,guzman13}. 
The origin of CH$_{3}$OH and H$_{2}$CO is, however, not yet fully understood.  
Pure gas-phase chemical models cannot always reproduce their observed abundances 
\citep{garrod06,geppert06,guzman11,guzman13}, 
and 
solid-phase chemistry, either on the surface of dust grains or in the ice mantles that accrete on top of them,  
must therefore play a role in their formation. 

Methanol is present in the solid and gaseous phases of the ISM. 
Solid CH$_{3}$OH has been detected in interstellar ice in dense cores with an abundance of 5 \% - 12 \% relative to water \citep{boogert11} 
and also in the cold envelopes around high and low-mass protostars with abundances ranging from 1\% up to 30\% relative to water 
\citep{dartois99,gibb00,gibb04,boogert08,pontoppidan08,bottinelli10}. 
Its abundance in solar system comets is lower \citep[0.2 \% - 7 \%,][and references therein]{mumma11}. 
Gas-phase CH$_{3}$OH has been detected in 
dense cores and photodissociation-dominated regions (PDRs) with a typical abundance of $\sim$10$^{-9}$ relative to H$_{2}$ 
\citep[see, e.g.,][]{bergman11,guzman11,guzman13}. 
This abundance is usually higher in hot cores \citep[$\sim$10$^{-7}$, ][and references therein]{oberg14}. 
In dense cores and PDRs where thermal desorption is inhibited, gas-phase models fail to reproduce the abundance of CH$_{3}$OH. 
Therefore, formation of this species in 
the solid phase 
and subsequently nonthermal desorption into the gas phase needs to be invoked 
\citep[and ref. therein]{guzman11,guzman13}. 

It has been proposed that CH$_{3}$OH is formed by successive addition of H atoms to CO molecules on the surface of 
dust grains \citep{tielens97,watanabe02,fuchs09} or, alternatively, in the bulk of water-rich ices \citep[e.g., ][]{watanabe03}. 
However, \citet{hiraoka02,hiraoka05} have claimed that this formation pathway cannot be the major source of methanol in the ISM, 
since the reaction rates are too slow. 
\citet{watanabe07} acknowledge that the contribution of photolysis mechanisms (see below) could not be negligible in molecular clouds. 
Therefore, energetic processing of CO molecules in a water-ice matrix without directly involving any addition of H atoms has been proposed as an 
alternative source of CH$_{3}$OH.  
For example, photolysis of H$_{2}$O:CO ice analogs \citep{schutte96}, 
proton bombardment of H$_{2}$O:CO ices \citep{hudson99},  
or electron irradiation of H$_{2}$O:CO:H$_{2}$O layered ices \citep{yamamoto04}.

Equivalent formation pathways using H$_{2}$O:CH$_{4}$ ice analogs have
also been proposed for the 
proton bombardment \citep{moore98} and the 
electron irradiation mechanisms \citep{wada06}, 
suggesting 
that the contribution of this mechanism for the formation of methanol was at least comparable to the H atom grain surface reactions. 
In the present work we have explored the photolysis of H$_{2}$O:CH$_{4}$ ice analogs. 
\citet{madzunkov10} propose a completely different formation route by collision of superthermal O atoms with CH$_{4}$ molecules in a methane ice 
with an overcoat of CO molecules. 

No UV-photon-induced desorption of CH$_{3}$OH was observed in \citet{gustavo15} during irradiation of a pure methanol ice, 
in contrast to what was previously reported in \citet{oberg09}. 
The low CH$_{3}$OH photodesorption of $\le$ 1.8 $\times$ 10$^{-4}$ molecules per incident photon 
found in UV irradiation experiments of pure methanol ices, 
probably due to the likely dissociation of CH$_{3}$OH molecules upon vacuum-UV-photon absorption, 
cannot account for the abundances of gas-phase methanol in cold regions, and formation followed by subsequent nonthermal desorption 
pathways must therefore be explored as an alternative. 

Formaldehyde has also been detected in both solid and gaseous phases. 
Solid H$_{2}$CO has not been detected yet in dense cores 
\citep{mumma11}, 
but it is present with an abundance of 1~\% - 6~\% relative to water in the cold envelopes around high- and low-mass protostars 
\citep{keane01,maret04,dartois05,boogert08}, 
while its abundance is a little lower in solar system comets 
\citep[0.1 - 1 \%,][and references therein]{mumma11}. 
Gas-phase H$_{2}$CO is also detected 
in dense cores and PDRs with a similar abundance to that of methanol \citep[see, e.g.,][]{guzman11,guzman13}  
and 
in hot cores,  
probably after thermal evaporation of ice mantles 
 \citep[e.g.,][]{wootten96,ceccarelli00,maret04,young04,araya07}. 
Although pure gas-phase models have been able to reproduce the abundance of H$_{2}$CO in dense cores, 
formation of this species on dust grains and subsequent (nonthermal) desorption is 
still  
needed to account for its abundance in low UV-field, illuminated PDRs   
\citep[for example, the Horsehead PDR, see][and references therein]{guzman11,guzman13}. 

Most of the solid-phase formation pathways proposed for H$_{2}$CO are shared with CH$_{3}$OH, 
since the former is usually an intermediate product in the formation of the latter. 
Formaldehyde can thus be produced by successive addition of H atoms to CO molecules 
\citep{tielens82,hiraoka94,watanabe02,fuchs09}. 
\citet{madzunkov09} report a similar formation route of H$_{2}$CO by collision of superthermal H atoms with CO molecules on a gold surface. 
Alternatively, photolysis \citep{schutte96}, proton bombardment \citep{hudson99}, and electron irradiation \citep{yamamoto04} of H$_{2}$O:CO ice analogs 
have reported the formation of H$_{2}$CO as an intermediate product in the formation of methanol. 
\citet{moore98} and \citet{wada06} studied the proton bombardment and electron irradiation of H$_{2}$O:CH$_{4}$ ice analogs, respectively. 
Photolysis of these binary ice samples is studied in the present work, as mentioned above. 
Formaldehyde can also be produced from 
UV or soft X-ray irradiation of pure methanol ices \citep[respectively]{oberg09,ciaravella10}. 
Collision of superthermal O atoms with CH$_{4}$ molecules also leads to the formation of H$_{2}$CO according to \citet{madzunkov10}.
 
Nonthermal desorption from pure formaldehyde ices has not yet been reported,  
but a behavior similar to that of pure methanol irradiation can be expected. Nonthermal desorption of molecules in cold regions can be induced by UV photons, cosmic rays, or exothermic chemical reactions 
\citep
[products formed in the surface of dust grains can immediately desorb due to the exothermicity of the reaction, ][]
{garrod07,hocuk14}. 
Experimental simulations dedicated to photoprocessing of ice analogs under astrophysically relevant conditions have revealed 
that UV-photon-induced desorption can take place through two different main mechanisms 
\citep[see also][and ref. therein]{martin15,gustavo15}. 

Photodesorption itself is an indirect desorption induced by electronic transitions (DIET) process where the energy provided by a UV photon to 
a molecule in the subsurface region of the ice is subsequently redistributed to surface molecules, which are able to break the intermolecular 
bonds and desorb into the gas phase 
\citep{rak95,oberg07,munozcaro10,fayolle11,bertin12,bertin13}. 
In this case, the desorbing molecules do not have to belong to the same species as the molecules absorbing the UV photons. 
\citet{fillion14} distinguish between DIET photodesorption, when both the absorbing and desorbing molecules belong to the same species, 
and indirect DIET photodesorption, when the molecule absorbing the UV photon is different from the molecule finally desorbing. 
In this paper the term DIET photodesorption is used to refer to both cases.  
This process therefore depends on both the composition and UV absorption spectra of the subsurface ice layers 
and the intermolecular bonds of molecules on the ice surface.  
Photochemidesorption, on the other hand, is the desorption of 
excited 
photofragments or photoproducts 
right after their formation on the surface of the ice 
\citep{martin15,gustavo15}, including non-DIET desorption processes previously reported in \citet{fayolle13} and \citet{fillion14}. 
This mechanism allows UV-photon-induced desorption of species that are not able to photodesorb significantly by irradiation of the pure ices 
\citep[e.g., CH$_{4}$ produced during CH$_{3}$OH irradiation,][]{gustavo15}. 

In the present work we have explored new UV-induced-formation pathways and subsequent desorption for CH$_{3}$OH and H$_{2}$CO.
On one hand, we have followed a top-down approach, studying the UV photodissociation of a pure ethanol ice   
\citep[solid C$_{2}$H$_{5}$OH has been proposed as one of the possible carriers of the 7.24 $\mu$m ice band, ][]{schutte99,oberg11}.
A bottom-up approach is represented by 
a more realistic binary H$_{2}$O:CH$_{4}$ ice analog, 
following the experiments already mentioned in this section. 

This paper is organized as follows. Section \ref{lab} describes the experimental simulations performed. 
The results are presented in Sect. \ref{resultados}
. 
The astrophysical implications are elucidated in Sect. \ref{imp},  
and the conclusions are summarized in Sect. \ref{conclusiones}. 

\section{Experimental}
\label{lab}

The experimental simulations have been performed using the InterStellar Astrochemistry Chamber (ISAC) at the Centro de Astrobiolog\'ia 
\citep{munozcaro10}. 
The ISAC setup consists in an ultra-high-vacuum (UHV) chamber with a base pressure of about 4 x 10$^{-11}$ mbar, similar to what is found 
in dense cloud interiors. 
Pure ethanol ices were grown by deposition of C$_{2}$H$_{5}$OH vapor onto a diamond window 
at 8 K, achieved by means of a closed-cycle helium cryostat.  
At this temperature, ices are deposited as amorphous solids. 
We used C$_{2}$H$_{5}$OH (liquid, 99.9\%) in this series of experiments. 
Binary 
ice analogs were grown in the interior of the chamber by simultaneous deposition of H$_{2}$O vapor and CH$_{4}$ gas onto the same substrate 
at 8K.   
The isotopolog 
$^{13}$CH$_{4}$ 
was used in similar experiments to confirm the results. 
The chemical components used in this series of experiments were H$_{2}$O (liquid, triply distilled), CH$_{4}$ (gas, Praxair 99.95\%), 
and $^{13}$CH$_{4}$ (gas, Cambridge Isotope Laboratories 99.9\%).

Ice samples were UV-irradiated using an F-type microwave-discharged hydrogen flow lamp (MDHL) from Opthos Instruments  
with a VUV-flux of $\approx$ 2 $\times$ 10$^{14}$ photons cm$^{-2}$ s$^{-1}$ at the sample position, measured by CO$_{2}$ $\to$ CO actinometry 
\citep{munozcaro10}. 
The size of the region irradiated by the lamp coincides with the size of the susbtrate. 
The spectrum of the MDHL is similar to the secondary UV field of dense cloud interiors calculated by \citet{gredel89} 
and also to the diffuse interstellar UV field \citep{jenniskens93,munozcaro03}, as well as to the far-UV field of emission/reflection 
nebulae \citep{france05}. 
It has been characterized previously by \citet{chen10,chen14}. 
More information can be found in \citet{gustavo14}, where a description of the vacuum-ultraviolet (VUV) spectrophotometer used in our setup to 
monitor the VUV flux is provided. 
The mean photon energy is 8.6 eV. 
A MgF$_{2}$ window is used as interface between the lamp and the chamber, leading to a cutoff at $\sim$114 nm (10.81 eV).

\begin{table}
\centering
\caption{IR feature used to calculate the column density of selected ice components. 
Frequencies and band strengths for pure ices at 10 K, except for C$_{2}$H$_{5}$OH (see text).}
\label{fuerzas}
\begin{tabular}{ccc}
\hline
\hline
Molecule&Frequency&Band strength\\
&(cm$^{-1}$)&(cm molec$^{-1}$)\\
\hline
C$_{2}$H$_{5}$OH&1044&7.3$\times 10^{-18}$ $^{a}$\\
H$_{2}$O&3280&2.0$\times 10^{-16}$ $^{b}$\\
CH$_{4}$&1304&6.4$\times 10^{-18}$ $^{c}$\\
%
CH$_{3}$OH&1025&1.8$\times 10^{-17}$ $^{d}$\\
H$_{2}$CO&1720&9.6$\times 10^{-18}$ $^{e}$\\
\hline
\end{tabular}
\begin{list}{}
\item $^{a}$ From \citet{moore98}\\
\item $^{b}$ From \citet{hagen1981}\\
\item $^{c}$ From \citet{dhendecourtallamandola86}. The same value was used as an approximation for the experiments with $^{13}$CH$_{4}$.\\
\item $^{d}$ From \citet{dhendecourtallamandola86}.\\
\item $^{e}$ From \citet{schutte96}\\
\end{list}
\end{table}

The ice samples were monitored 
by in situ Fourier-transform infrared (FTIR) transmittance spectroscopy 
after deposition and after every irradiation period, using a Bruker Vertex 70 spectrometer equipped with a deuterated triglycine sulfate detector 
(DTGS). 
The IR spectra were collected 
with a spectral resolution of 2 - 4 cm$^{-1}$ (most spectra were subsequently smoothed). 
Column densities of selected species in the ice were calculated from the IR spectra using the formula

\begin{equation}
N=\frac{1}{A}\int_{band}{\tau_{\nu} \ d\nu},
\end{equation}

\noindent where $N$ is the column density in molecules cm$^{-2}$, $\tau_{\nu}$ the optical depth of the absorption band, and $A$ the band strength in cm molecule$^{-1}$, 
as derived from laboratory experiments (Table \ref{fuerzas}). 
Band strengths in Table \ref{fuerzas} were measured for pure amorphous ices execpt for ethanol. 
The same values are usually adopted in ice mixtures, which introduces an uncertainty of about 20-30\% \citep{dhendecourtallamandola86}. 
In the case of ethanol, the band strength in Table \ref{fuerzas} was measured in a mixture with amorphous water ice at T \textless 20 K 
\citep{moore98}. Using this value in a pure amorphous ethanol ice introduces a similar uncertainty than that mentioned above.  

\begin{figure}
\centering
 \includegraphics[width=7.cm]{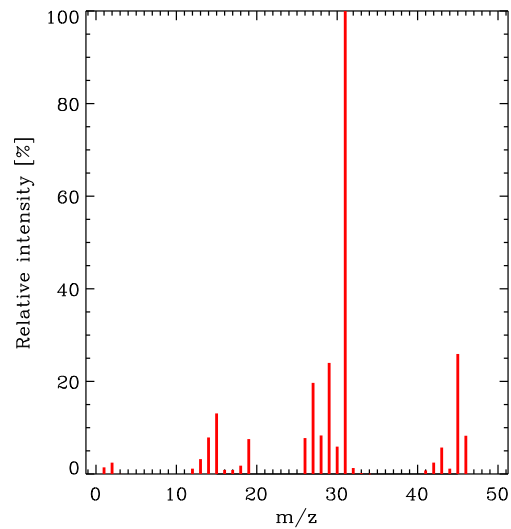}
 \caption{Mass spectrum of ethanol in our setup.} 
\label{espectro_etanol}
\end{figure}

\begin{figure}
\centering
 \includegraphics[width=7.cm]{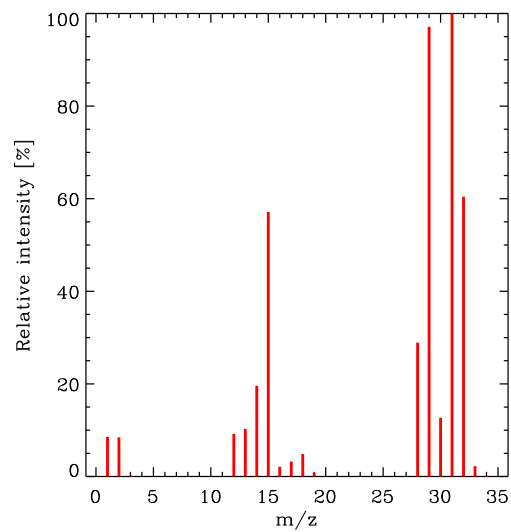}
 \caption{Mass spectrum of methanol in our setup.} 
\label{espectro_metanol}
\end{figure}

A 
Pfeiffer Prisma quadrupole mass spectrometer (QMS) of a mass spectral range 
from 1 to 200 amu with a Channeltron detector 
was used to detect molecules in the gas phase. 
The QMS ionizes gas-phase molecules with $\sim$ 70 eV electron bombardment, 
leading to fragmentation of the molecules with a given pattern. 
The species 
potentially desorbing into the phase 
were preferably 
monitored through their main mass fragment,  
which does
not always coincide with the molecular ion. 
In some cases the main mass fragment of a given species was common to a mass fragment of another species 
(see, for example, Figs. \ref{espectro_etanol} and \ref{espectro_metanol} for the mass spectrum of 
ethanol and methanol, respectively, in our setup).  
When possible, alternative mass fragments were used in those cases (see Table \ref{fragmentos}). 
Fragments in Table \ref{fragmentos} were also used in the experiments with the isotopically labeled molecules,  
but mass fragments with $^{13}$C were 1 amu higher.  
There were two exceptions: $^{13}$CH$_{4}$ was monitored through the mass fragment m/z=15, which corresponds to the $^{13}$CH$_{2}^{+}$ fragment, 
and $^{13}$CH$_{3}$OH was monitored through the mass fragment m/z=33, which corresponds to the molecular ion. 

\begin{table*}
\centering
\caption{Mass fragments used to monitor potentially desorbing species into the gas phase.}
\label{fragmentos}
\begin{tabular}{ccccl}
\hline
\hline
Molecule&Molecular ion&Main mass fragment&Mass fragment used&Notes\\
\hline
C$_{2}$H$_{5}$OH&45&31&45&m/z=31 was common to CH$_{3}$OH\\
H$_{2}$O&18&18&18&\\
CH$_{4}$&16&16&15&m/z=16 was common to H$_{2}$O\\
&&&&Contribution from CH$_{3}$OH molecules was negligible\\
CH$_{3}$OH&32&31&31&CH$_{3}$OH could not be unambiguously detected in the\\
&&&&gas phase during C$_{2}$H$_{5}$OH irradiation (see text)\\
H$_{2}$CO&30&29&29&m/z=30 was also used (see text)\\
CH$_{3}$CHO&44&29&43&m/z=29 was common to H$_{2}$CO\\
&&&&m/z=44 was common to CO$_{2}$\\
&&&&CH$_{3}$CHO was not detected in the gas phase during \\
&&&&irradiation in both series of experiments\\
C$_{2}$H$_{6}$&30&28&27&m/z=28 was common to CO\\
&&&&m/z=30 was common to H$_{2}$CO\\
CO&28&28&28&Contribution from other species was negligible\\
CO$_{2}$&44&44&44&CO$_{2}$ was not detected in the gas phase during\\
&&&&irradiation in both series of experiments\\
\hline
\end{tabular}
\end{table*}


The targeted molecules of this work were CH$_{3}$OH and H$_{2}$CO. 
Methanol was monitored through its main mass fragment, m/z=31. 
However, in the irradiation experiments of pure C$_{2}$H$_{5}$OH ice, m/z=31 was common to the main mass fragment of the ethanol molecules.  
The rest of mass fragments with a reasonable abundance, according to Fig. \ref{espectro_metanol} (m/z=29, m/z=32, m/z=15, m/z=28, and m/z=30), 
were also common to H$_{2}$CO, O$_{2}$, CH$_{4}$, CO, and H$_{2}$CO molecules, respectively.
Therefore, CH$_{3}$OH could not be unambiguously detected in the gas phase in that series of experiments. 
The main mass fragment of H$_{2}$CO is m/z=29, which was common to 
ethanol, methanol, and ethane (see below).  
Ethanol molecules could contribute in the first series of experiments with up to a 20\% of the total m/z=29 signal, 
according to the mass spectrum of this molecule in Fig. 
\ref{espectro_metanol}, since the level of the m/z=31 signal (likely coming from ethanol, as explained below) was found to be comparable to 
the m/z=29 mass fragment. 
Contribution from methanol molecules during the first series of experiments was not likely, 
since the behavior of the m/z=31 signal follows that of the m/z=45 signal, and it was therefore coming from ethanol molecules. 
Methanol was not detected in the gas phase in the second series of experiments. 
Contribution from ethane molecules could not be completely discarded during the first series of experiments, though. 
Although the second most abundant fragment (m/z=30) was used to double-check the detection formaldehyde molecules in the gas phase, 
contribution of ethane molecules to this fragment could not be discarded either in the first series of experiments.  

Calibration of our QMS allows conversion from integrated ion currents into photon-induced desorbed column densities, using the equation

\begin{equation}
N (mol) = \frac{A (m/z)}{k_{CO}} \cdot \frac{\sigma^{+} (CO)}{\sigma^{+} (mol)} \cdot \frac{I_{F} (CO^{+})}{I_{F} (z)} 
\cdot \frac{F_{F} (28)}{F_{F} (m)} \cdot \frac{S (28)}{S (m/z)} 
\label{eqmscal}
,\end{equation}

\noindent where 
$N (mol)$ is the total number of photon-induced desorbed molecules cm$^{-2}$, 
$A (m/z)$ the integrated area below the QMS signal of a given mass fragment $m/z$ during photon-induced desorption, 
$k_{CO}$ is the proportionality constant between the integrated ion current and the column density of desorbed molecules 
in the case of a pure CO ice, as shown by

\begin{equation}
k_{CO} = \frac{A (28)}{N (CO)} = k_{QMS} \cdot \sigma^{+} (CO) \cdot I_{F} (CO^{+}) \cdot F_{F} (28) \cdot S (28) 
\label{eqmsco}
,\end{equation}

\noindent with $k_{QMS}$ the proportionality constant independent of the species.  
The constant $k_{CO}$ is regularly calculated with pure CO ice irradiation experiments.   
Parameter $\sigma^{+} (mol)$ is the ionization cross section for the first ionization of the species of interest 
and the incident electron energy of the mass spectrometer; 
$I_{F} (z)$ is the ionization factor, that is, the fraction of ionized molecules 
with charge $z$;
$F_{F} (m)$ the fragmentation factor, that is, the fraction of molecules 
of the isotopolog of interest 
leading to a fragment of mass $m$ in the mass spectrometer; and
$S (m/z)$ the sensitivity of the QMS to the mass fragment $(m/z)$. 
In practice, we work with the product $k^{*}_{QMS} \cdot S (m/z)$ 
($k^{*}_{QMS}$ indicates that pressure units are used instead of column density units), 
since the ratio  $S (m/z)$/$S (28)$ is the same as the ratio $k^{*}_{QMS} \cdot S (m/z)$/$k^{*}_{QMS} \cdot S (28)$. 
Sensitivity of the QMS is also probed regularly, using noble gases. 
Calibration of the QMS is detailed in \citet{martin15}. 

Equations \ref{eqmscal} and \ref{eqmsco} assume that the pumping speed in the ISAC setup is the same for all molecules, so that $k_{QMS}$ does 
not depend on the species \citep[see ][]{martin15}. 
In fact, the pumping speed depends on the molecular mass and, to a lesser extent, on the molecular structure \citep{kaiser95}. 
Constant $k_{QMS}$ in equation \ref{eqmsco} thus corresponds to the pumping speed of CO molecules. 
Therefore, $N (mol)$ calculated with equation \ref{eqmscal} is only valid if the pumping speed of a specific species is the same as for CO. 
Taking the different pumping speeds into account, the real number of photon-induced desorbed molecules would be 

\begin{equation}
N^{real} (mol) = N^{calc} (mol) \cdot S_{rel} (mol) 
\label{eqmscorr}
,\end{equation}

\noindent with $N^{calc} (mol)$ the column density calculated with eq. \ref{eqmscal}, and $S_{rel} (mol)$ the relative 
pumping speed with respect to the CO molecules. 
According to the manufacturer of the pumping devices used in the ISAC setup, the relative pumping speed with respect to CO 
of a species with molecular mass $M (mol)$ is 

\begin{equation}
S_{rel} (mol) = 1.258 - 9.2 \cdot 10^{-3} \cdot M (mol)
\label{eqmspump}
.\end{equation}

\noindent Photon-induced desorption yields can be subsequently calculated by dividing $N^{real} (mol)$ by the fluence (the product of the 
VUV-flux and the irradiation time).  

A 
small rise in the signal of several mass fragments was detected every time the UV lamp was switched on.  
This effect has been previously reported in other works \citep[see, e.g., ][]{loeffler05}, but it was not reproducible so was 
difficult to quantify. 
When comparable to the rise produced by photon-induced desorption of molecules from the ice during the first series of experiments, 
calculated desorption yields were considered upper limits. 

At the end of the experimental simulations, ice samples were warmed up to room temperature using a LakeShore Model 331 temperature controller, 
until a complete sublimation was attained. 
A silicon diode temperature sensor located close to the ice substrate was used, reaching a sensitivity of about 0.1 K. The 
IR spectra of the solid sample were collected every five minutes during warm-up. 
At the same time, all the desorbing species, including the components of the ice analogs and the products of the photochemical reactions, 
were detected by the QMS mentioned above. 

\begin{table*}
\centering
\caption{UV photoprocessing experiments of pure C$_{2}$H$_{5}$OH ices.}
\label{exp_et}
\begin{tabular}{cccccc}
\hline
\hline
Experiment&N$_{initial}$(C$_{2}$H$_{5}$OH)&Fluence&N$_{final}$(C$_{2}$H$_{5}$OH)&N$_{final}$(CH$_{3}$OH)&N$_{final}$(H$_{2}$CO)$^{a}$\\
&$\times$10$^{15}$molecules cm$^{-2}$&$\times$10$^{17}$photons cm$^{-2}$&\multicolumn{3}{c}{$\times$10$^{15}$molecules cm$^{-2}$}\\
\hline
1&204.7&7.2&39.8&0.0&$\le$90.1\\
2&185.6&7.2&38.3&0.0&$\le$85.5\\
\hline
\end{tabular}
\begin{list}{}
\item $^{a}$ This is an upper limit since the H$_{2}$CO IR band used is blended with two more bands corresponding to acetaldehyde and water.\\
\end{list}
\end{table*}

\section{Experimental results and discussion}
\label{resultados}
\subsection{UV photoprocessing of a pure C$_{2}$H$_{5}$OH ice}
\label{et}

Table \ref{exp_et} presents the two experiments performed involving UV photoprocessing of a pure C$_{2}$H$_{5}$OH ice. 
The production of new species due to the energetic processing of the ice samples is studied in Sect. \ref{chem_et} by means of 
IR spectroscopy and mass spectrometry (during temperature-programmed desorption (TPD) of the ice sample at the end of the experiment). 
Photon-induced desorption of 
the photoproducts is studied in Sect. \ref{des_et} by means of mass spectrometry. 

\begin{figure*}
\centering
 \includegraphics[width=13.cm]{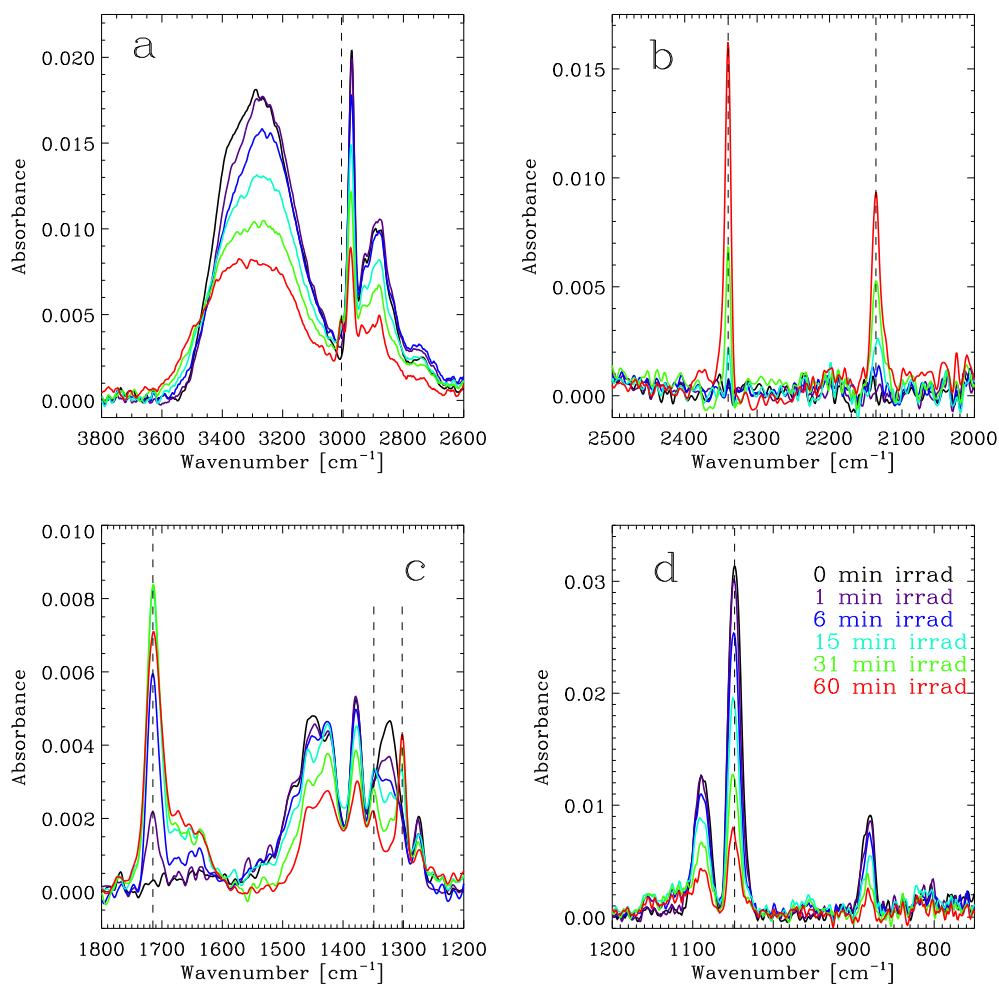}
 \caption{Evolution of the IR spectrum of the ice sample during UV photoprocessing in experiment 2 (see Table \ref{exp_et}) conveniently 
 split in four regions. Results were similar in experiment 1. 
\textbf{a)} A new feature appears at 3005 cm$^{-1}$ (dashed line) between the O-H and C-H stretching bands of C$_{2}$H$_{5}$OH, 
which is attributed to the formation of CH$_{4}$. This species also displays a more intense band near 1302 cm$^{-1}$, shown in panel c. 
\textbf{b)} Formation of CO$_{2}$ and CO leads to the appearance of two C-O stretching bands at 2340 cm$^{-1}$ and 2136 cm$^{-1}$ (dashed lines), 
respectively. 
\textbf{c)} Dashed lines at 1715 cm$^{-1}$, 1349 cm$^{-1}$, and 1302 cm$^{-1}$ indicate new features due 
to the formation of photoproducts (probably H$_{2}$CO, CH$_{3}$CHO, C$_{2}$H$_{6}$, and CH$_{4}$; see text). 
A broad band on the red side of the 1715 cm$^{-1}$ peak may indicate the presence of H$_{2}$O. 
\textbf{d)} The intensity of the C-O stretching band of C$_{2}$H$_{5}$OH at 1048 cm$^{-1}$ (dashed line) decreases as the ice is processed.} 
\label{IRetanol}
\end{figure*}

\subsubsection{Photon-induced chemistry of pure C$_{2}$H$_{5}$OH ice}
\label{chem_et}

\begin{figure*}
\centering
 \includegraphics[width=13.cm]{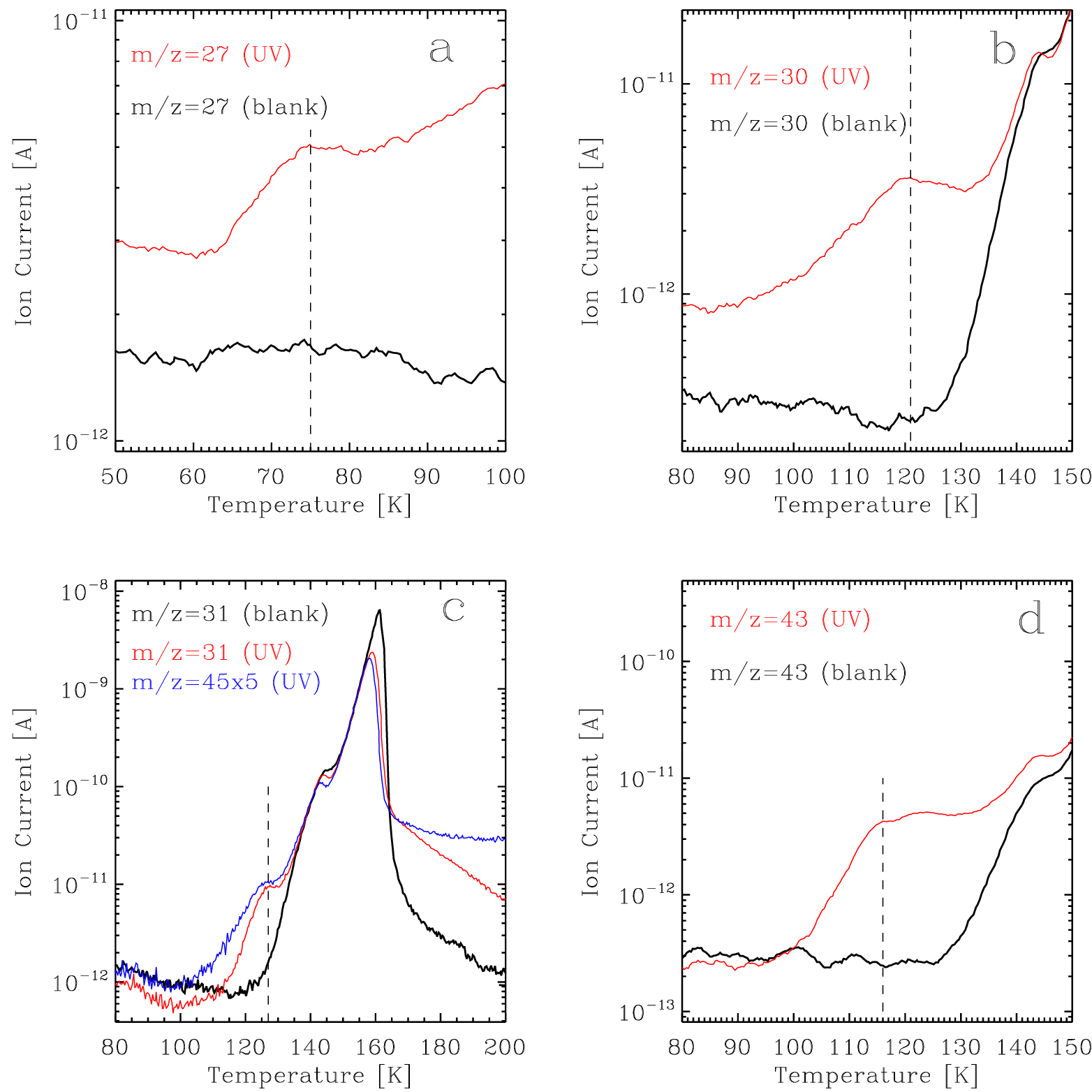}
 \caption{TPD curves displaying the thermal desorption of photoproducts through selected mass fragments at the end of experiment 2 
 (see Table \ref{exp_et}) in red, compared to the blank experiment with no irradiation (black). 
\textbf{a)} TPD curves of the m/z=27 mass fragment, corresponding to the C$_{2}$H$_{3}^{+}$ fragment of C$_{2}$H$_{6}$.  
A desorption peak at $\sim$75 K (dashed line) confirms the presence of ethane in the processed ethanol ice 
\citep[desorption of a pure ethane ice takes place above 60 K according to][]{oberg09}.
\textbf{b)} TPD curves of the m/z=30 mass fragment, corresponding to the molecular ion of H$_{2}$CO. 
The desorption peaking at $\sim$121 K (dashed line) confirms the formation of formaldehyde during UV photoprocessing of a pure ethanol ice (see text).
\textbf{c)} TPD curves of the m/z=31 mass fragment corresponding to the CH$_{3}$O$^{+}$ fragment of CH$_{3}$OH. 
Since no excess in the desorption is detected at $\sim$145 K compared to the blank, formation of methanol during UV photoprocessing of a pure 
ethanol ice is negligible. 
Desorption at $\sim$127 K (dashed line) is probably due to an earlier ethanol desorption triggered by less polar and more volatile 
components in the ice matrix that were not present in the blank experiment (mainly formaldehyde and acetaldehyde), since it is shared with 
the m/z=45 mass fragment (blue). 
\textbf{d)} TPD curves of the m/z=43 mass fragment, corresponding to the CH$_{3}$CO$^{+}$ fragment of CH$_{3}$CHO. 
The desorption peaking at $\sim$116 K (dashed line) confirms the formation of acetaldehyde during UV photoprocessing of a pure ethanol ice (see text).
} 
\label{TPDetanol}
\end{figure*}

An IR spectrum of the ice sample was collected after every irradiation interval to monitor the ice composition in the two experiments. 
Figure \ref{IRetanol} shows the evolution of the mid-IR spectrum (between 3800 cm$^{-1}$ and 750 cm$^{-1}$) 
of the ice sample in experiment 2. Similar results were found during experiment 1. 

The IR spectrum of the ice sample is dominated by the C$_{2}$H$_{5}$OH features between 3600 cm$^{-1}$ and 2650 cm$^{-1}$ 
(Fig. \ref{IRetanol}a), 1580 cm$^{-1}$ and 1230 cm$^{-1}$ (Fig. \ref{IRetanol}c), and 1120 cm$^{-1}$ and 860 cm$^{-1}$ (Fig. \ref{IRetanol}d). 
New IR features appearing as the result of the UV photoprocessing of the ice samples were assigned to the formation of photoproducts. 
These assignments could be subsequently confirmed by monitoring the corresponding mass fragment during the TPD carried out at the 
end of the experiments. Figure \ref{TPDetanol} shows TPD spectra of selected mass fragments at the end of experiment 2. 
Results were similar during experiment 1.

The C$_{2}$H$_{5}$OH column density was best monitored with the 
IR band peaking at 1048 cm$^{-1}$ 
shown in Fig. \ref{IRetanol}d. 
The intensity of this band gradually decreases with continuing irradiation, 
and $\sim$80\% of the initial ethanol was either photodissociated or, to a lesser extent, photodesorbed after a total fluence of 
7.2 $\times$10$^{17}$photons cm$^{-2}$ in the two experiments (see Table \ref{exp_et}). 
A similar behavior is seen for the other C$_{2}$H$_{5}$OH features in Figs. \ref{IRetanol}a, \ref{IRetanol}c, and \ref{IRetanol}d. 

Photodissociation of C$_{2}$H$_{5}$OH triggered a complex photochemical network whose complete study is beyond the scope of this work. 
%
No evidence of CH$_{3}$OH formation was found during irradiation of the ice samples. In particular, the C-O stretching mode of CH$_{3}$OH 
that should appear at $\sim$1016 cm$^{-1}$ in a relatively clean region of the spectrum is not detected in Fig. \ref{IRetanol}d. 
Formation of CH$_{3}$OH in an extent below the sensitivity limit of our FTIR spectrometer is rejected since no desorption peak is detected 
with the more sensitive QMS 
for the m/z=31 mass fragment at the desorption temperature of methanol \citep[$\sim$145 K according to][see Fig. \ref{TPDetanol}c]{martin14}.
On the other hand, a new IR band is clearly detected at 1715 cm$^{-1}$ in Fig. \ref{IRetanol}c,  
due to the C=O stretching mode of a carbonyl group, reaching its maximum intensity after 31 minutes of irradiation. 
Both H$_{2}$CO and CH$_{3}$CHO contribute to this band, since desorption peaks 
at $\sim$121 K and $\sim$116 K were detected for the mass fragments m/z=30 (Fig. \ref{TPDetanol}b) and m/z=43 (Fig. \ref{TPDetanol}d), 
respectively, confirming the presence of these species in the processed ice   
\citep[desorption temperatures of pure H$_{2}$CO and CH$_{3}$CHO ices are $\sim$112 K and $\sim$105 K, respectively, according to][]
{noble12,oberg09}. 
An upper limit to the column density of H$_{2}$CO in the ice is shown in Table \ref{exp_et}. 
Up to $45\%$ of the initial ethanol was converted into formaldehyde at the end of the experiments. 

A broad band on the red side of the C=O stretching mode of carbonyl groups is due to blending with the water O-H bending mode. 
%
Other four carbon-bearing species were formed during UV photoprocessing of the pure ethanol ice. 
The C-H stretching and the deformation modes of CH$_{4}$ are clearly detected at 3005 cm$^{-1}$ and 1302 cm$^{-1}$ after 15 minutes of irradiation (see Figs. \ref{IRetanol}a and \ref{IRetanol}c, respectively). 
The deformation mode of C$_{2}$H$_{6}$ is also detected at 1349 cm$^{-1}$ (Fig. \ref{IRetanol}c) after six minutes of irradiation. 
The presence of ethane in the ice sample was confirmed during the TPD at the end of the experiments (see Fig. \ref{TPDetanol}a). 
Finally, the C-O stretching band of CO is clearly detected at 2136 cm$^{-1}$, 
while the C-O stretching band of CO$_{2}$ molecules is observed at 2340 cm$^{-1}$ 
(Fig. \ref{IRetanol}b). 
%

\begin{figure*}
\centering
 \includegraphics[width=18.cm]{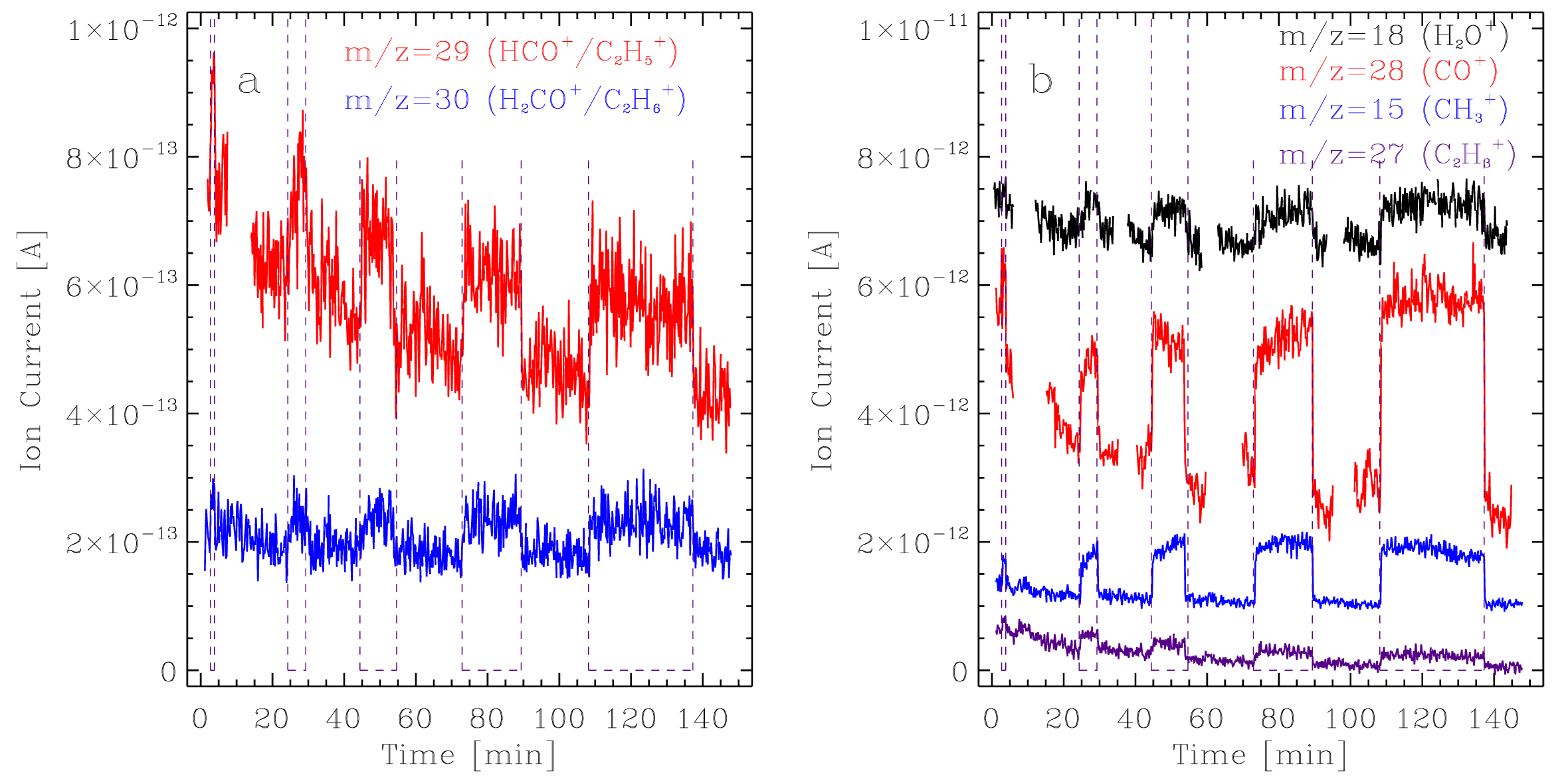}
 \caption{Photon-induced desorption of several photoproducts during irradiation of a pure ethanol ice in experiment 2, see Table \ref{exp_et}.  
Results were similar in experiment 1. 
Irradiation intervals are indicated with vertical dashed lines.  
Signal due to rotation of the sample holder between irradiation intervals were omitted. 
\textbf{a)} Phochemidesorption of formaldehyde/ethane (see text) observed with mass fragments m/z=29 and m/z=30. 
\textbf{b)} Photochemidesorption of water and ethane observed with mass fragments m/z=18 and m/z=27, respectively. 
DIET photodesorption was detected for CO and CH$_{4}$ (m/z=28 and m/z=15, respectively). 
Ion currents of different mass fragments are offset for clarity.}
\label{irradiacion_etanol}
\end{figure*}

\subsubsection{Photon-induced desorption from pure C$_{2}$H$_{5}$OH ice}
\label{des_et}

A QMS was used during UV photoprocessing of the ice samples to monitor the photon-induced desorption of the photoproducts detected in 
Sect. \ref{chem_et} using the mass fragments selected in Sect. \ref{lab}. 
Results are shown in Fig. \ref{irradiacion_etanol}.

Photon-induced desorbing species presented a rise in the QMS ion current of their corresponding mass fragments during irradiation. 
As mentioned in Sect. \ref{lab}, calibration of the QMS allowed us to obtain desorption yields of the photon-induced desorbing species 
from the integrated ion currents. 
Parameters used in Eq. \ref{eqmscal} for the photoproducts in experiments 1 and 2 are shown in Table \ref{param}. 
Proportionality constant $k_{CO}$ was 1.74 x 10$^{-10}$ A min ML$^{-1}$ for both experiments. 
Photon-induced desorption yields of all photoproducts in pure ethanol ice irradiation experiments 
(except for CH$_{3}$CHO and CO$_{2}$, whose desorption, if taking place, was negligible) 
are presented in Table \ref{ypdet} for experiment 2, averaged for each irradiation period. Results were similar in experiment 1. 

\begin{table*}
\centering
\caption{Values used in Eq. \ref{eqmscal} to convert integrated QMS signals for the mass fragments 
into column densities of desorbing molecules.}
\label{param}
\begin{tabular}{cccccc}
\hline
\hline
Factor&H$_{2}$CO&C$_{2}$H$_{6}$&H$_{2}$O&CH$_{4}$&CO\\
\hline
$\sigma^{+} (mol)$ ($\AA^{2}$)$^{a}$&4.140&6.422&2.275&3.524&2.516\\
fragment&H$_{2}$CO$^{+}$&C$_{2}$H$_{6}^{+}$&H$_{2}$O$^{+}$&CH$_{3}^{+}$&CO$^{+}$\\
$m/z$&30&30&18&15&28\\
$I_{F} (z)$&1$^{b}$&1$^{b}$&1$^{b}$&1$^{b}$&1$^{b}$\\
$F_{F} (m)$&0.319$^{a}$&0.121$^{c}$&0.806$^{c}$&0.406$^{c}$&0.949$^{c}$\\
$k^{*}_{QMS} \cdot S (m/z)$ (A mbar$^{-1}$ $\AA^{-2}$)$^{d}$ &1.77 x 10$^{15}$&1.77 x 10$^{15}$&3.86 x 10$^{15}$&4.69 x 10$^{15}$&2.02 x 10$^{15}$\\
$k^{*}_{QMS} \cdot S (m/z)$ (A mbar$^{-1}$ $\AA^{-2}$)$^{e}$&9.30 x 10$^{14}$&9.30 x 10$^{14}$&1.69 x 10$^{15}$&1.97 x 10$^{15}$&1.03 x 10$^{15}$\\ 
$k^{*}_{QMS} \cdot S (m/z)$ (A mbar$^{-1}$ $\AA^{-2}$)$^{f}$&1.29 x 10$^{15}$&1.29 x 10$^{15}$&2.50 x 10$^{15}$&2.94 x 10$^{15}$&1.44 x 10$^{15}$\\ 
$S_{rel} (mol)^{g}$&0.982&0.982&1.092&1.111&1.000\\
\hline
\end{tabular}

\begin{list}{}
\item $^{a}$ Extracted from the online database of the National Institute of Standard and Technologies (NIST).\\
\item $^{b}$ A value of 1 was adopted, assuming that no double ionization of the molecules takes place.\\
\item $^{c}$ Extracted from the mass spectral library of the QMS software.\\
\item $^{d}$ Calculated for our QMS and valid for experiments 1, 2, 6, 9, and 10 (see Sect. \ref{lab}).\\
\item $^{e}$ For experiments 3 and 7 (see Sect. \ref{lab}).\\
\item $^{f}$ For experiments 4, 5, and 8 (see Sect. \ref{lab}).\\
\item $^{g}$ Calculated with Eq. \ref{eqmscorr}.\\
\end{list}

\end{table*}

\begin{table*}
\centering
\caption{Evolution of the photon-induced desorption yields during UV irradiation of a pure C$_{2}$H$_{5}$OH ice.}
\label{ypdet}
\begin{tabular}{ccccccc}
\hline
\hline
Irradiation period&Fluence$^{a}$&$Y_{pd}$ (H$_{2}$CO)$^{b,c}$&$Y_{pd}$ (C$_{2}$H$_{6}$)$^{b,c}$&$Y_{pd}$ (H$_{2}$O)$^{b,c}$&
$Y_{pd}$ (CH$_{4}$)$^{b}$&$Y_{pd}$ (CO)$^{b}$\\
(min)&(photons cm$^{-2}$)&10$^{-5}$ (molecules/incident photon)&
\multicolumn{3}{c}{10$^{-4}$ (molecules/incident photon)}
&10$^{-3}$ (molecules/incident photon)\\
\hline
0 - 1&1.2 x 10$^{16}$&$\le$7.6&$\le$1.3&$\le$1.1&1.8&0.6\\
1 - 6&7.2 x 10$^{16}$&$\le$5.9&$\le$1.0&$\le$2.0&2.3&0.7\\
6 - 15&1.8 x 10$^{17}$&$\le$5.5&$\le$1.0&$\le$2.7&3.1&1.0\\
15 - 31&3.7 x 10$^{17}$&$\le$6.4&$\le$1.1&$\le$2.4&3.4&1.1\\
31 - 60&7.2 x 10$^{17}$&$\le$5.8&$\le$1.0&$\le$2.7&3.2&1.5\\
\hline
\end{tabular}
\begin{list}{}
\item $^{a}$ Total fluence at the end of the irradiation period.\\ 
\item $^{b}$ Averaged for each irradiation period in experiment 2. Results were similar in experiment 1.
Photon-induced desorption yield values could be different by a factor 2 due to the uncertainties in all the parameters 
of Eq. \ref{eqmscal}. 
\item $^{c}$ We consider these values upper limits due to possible contribution from other species to the signal of the selected mass fragment  
and/or from the small rise experienced by the ion currents when the UV lamp is switched on (see Sect. \ref{lab}). 
\end{list}
\end{table*}

Evolution of the photodesorption yield with fluence is related to the two main photon-induced desorption mechanisms presented 
in Sect. \ref{intro}. 
Photoproducts that are able to desorb through the process called indirect desorption induced by electronic transitions (DIET) 
increase their photodesorption yield (or in other words, their QMS ion current during irradiation) with fluence, since molecules 
previously formed and 
accumulated in the bulk of the ice are later available for desorption when a photon is absorbed by a nearby molecule 
(indirect DIET photodesorption is also included).  
On the other hand, photochemidesorbing-only species 
do not increase their 
photodesorption yield (or ion current) with fluence, 
since only molecules formed on the surface of the ice are able to desorb right after their formation, 
and molecules accumulated in the bulk of the ice are not able to desorb later on during irradiation. 
In the case of molecules able to both photochemidesorb and desorb through the DIET mechanisms, the effect of the former will be negligible 
compared to the latter since the number of molecules at the surface of the ice is much lower than in the bulk \citep[see also][] {fillion14}. 

Photon-induced desorption of methanol was discarded since the behavior of the m/z=31 signal follows that of the m/z=45 signal, 
and it was therefore coming from ethanol molecules.
On the other hand, ion currents of mass fragments m/z=29 and m/z=30 in Fig. \ref{irradiacion_etanol}a present rather constant rises during 
UV irradiation of the ice sample, suggesting photochemidesorption. 
As explained in Sect. \ref{lab}, either formaldehyde and/or ethane molecules could be responsible for the detected mass fragments, 
since a similar rise is detected for the ethane mass fragment m/z=27 in Fig. \ref{irradiacion_etanol}b.  
This means that one out of two, or even both species could be photochemidesorbing to the gas phase during irradiation of the pure ethanol ice. 
Therefore, we used the integrated ion currents of mass fragment m/z=30 and the parameters shown in Table \ref{param} for both species 
to extract upper limits to their photon-induced desorption yields (see Table \ref{ypdet}). 
The average photon-induced desorption yield would be of $\sim$6.2 x 10$^{-5}$ molecules/incident photon if the 
m/z=30 signal was due only to formaldehyde molecules, and of $\sim$1.1 x 10$^{-4}$ molecules/incident photon if
only ethane molecules contributed to the signal. 
These values should be considered upper limits due to the possible contribution from the small rise experienced by the ion current when the 
UV lamp was switched on during this series of experiments (see Sect. \ref{lab}). 
In both cases the photon-induced desorption yields remain almost constant with fluence, as expected for photochemidesorbing species. 


Figure \ref{irradiacion_etanol}b shows the evolution during photoprocessing of a pure ethanol ice for the mass fragments corresponding to 
other photoproducts detected in Sect. \ref{chem_et}.   
Apart from the m/z=27 mass fragment corresponding to ethane molecules commented above, 
the ion current of mass fragments m/z=15 (CH$_{3}^{+}$), m/z=18 (H$_{2}$O$^{+}$), and m/z=28 (CO$^{+}$) were found to increase from one 
irradiation interval to the next, indicating that the photoproduced methane, water, and carbon monoxide, respectively, 
are able to desorb through the DIET mechanism during irradiation of a pure ethanol ice. 
In the case of CH$_{4}$, this behavior is different from the one experienced by the molecules 
photoproduced in a pure methanol ice, which were able to photochemidesorb but not to desorb through the DIET mechanism \citep{gustavo15}. 

In pure methane ice, CH$_{4}$ molecules are not able to desorb, 
probably because methane molecules are readily photodissociated by the VUV photons, and they cannot trigger the 
DIET photodesorption mechanism. 
Therefore, in a methanol or ethanol ice, the DIET photodesorption process of methane molecules would be triggered more likely by 
the absorbtion of the VUV photons by molecules of other species in the subsurface layers of the ice (indirect DIET). 
The different behavior in these two ice irradiation experiments indicates that the energy redistributed to methane molecules in the surface 
of the ice is enough to break the intermolecular bonds in the case of pure ethanol ice irradiation experiments, but not during irradiation 
of a pure methanol ice. 
This may be due to a higher intermolecular binding energy of methane molecules to their neighbors in the methanol 
ice irradiation experiments or, alternatively, to a higher energy redistributed by the absorbing molecules to the surface in the ethanol 
ice experiments. 

The photodesorption yield of water and methane molecules reached a constant value after approximately 15 minutes of irradiation, 
of $\le$2.6 x 10$^{-4}$ molecules/incident photon, 
and $\sim$3.3 x 10$^{-4}$ molecules/incident photon, respectively (see Table \ref{ypdet}). 
The photodesorption yield of CO molecules increased with fluence (Table \ref{ypdet}) during the full experiment. 

\begin{table*}
\centering
\caption{UV photoprocessing experiments of H$_{2}$O:CH$_{4}$ ice analogs.}
\label{exp_bin}
\begin{tabular}{cccccccc}
\hline
\hline
Experiment&N$_{initial}$(H$_{2}$O)&N$_{initial}$(CH$_{4}$)&Fluence&N$_{final}$(H$_{2}$O)&N$_{final}$(CH$_{4}$)&
N$_{final}$(CH$_{3}$OH)&N$_{final}$(H$_{2}$CO)$^{a}$\\
&\multicolumn{2}{c}{$\times$10$^{15}$molecules cm$^{-2}$}
&$\times$10$^{17}$photons cm$^{-2}$
&\multicolumn{4}{c}{$\times$10$^{15}$molecules cm$^{-2}$}\\
\hline
3&426.7&119.2&7.2&376.0&62.0&7.2&...\\
4&586.2&229.6&7.8&520.9&173.8&7.8&...\\
5&898.2&216.3&10.8&878.2&153.7&4.9&...\\
6&560.1&161.3&7.2&500.1&101.2&4.2&...\\
\hline
7$^{b}$&506.6&150.6&7.2&462.4&89.0&0.0&...\\
8$^{b}$&721.9&263.0&10.8&665.4&165.3&0.4&...\\
9$^{b}$&372.9&160.2&7.2&335.0&90.6&1.8&...\\
10$^{b}$&523.7&146.8&6.0&...$^{c}$&...$^{c}$&...$^{c}$&...$^{c}$\\
\hline
\end{tabular}
\begin{list}{}
\item $^{a}$ Formaldehyde was barely detected by means of IR spectroscopy, and no quantification was possible. Formation of H$_{2}$CO was confirmed 
during the TPD.\\
\item $^{b}$ $^{13}$CH$_{4}$ was used instead of CH$_{4}$ in these experiments.\\
\item $^{c}$ No IR spectrum was collected at the end of this experiment, since it focused only on measuring QMS signals during irradiation. 
(see Fig. \ref{irradiacion_binariais})\\
\end{list}
\end{table*}

\subsection{UV photoprocessing of a H$_{2}$O:CH$_{4}$ ice analog}
\label{bin}

The second series of experiments involved the UV photoprocessing of a more realistic, water-rich ice analog. 
A binary H$_{2}$O:CH$_{4}$ ice analog was used in experiments 3 to 10 described in Table \ref{exp_bin}. 
The initial ice composition in this series of experiments was held, approximately, to 75\% of water and 25\% of methane.  
The irradiation time changed from one experiment to the next so that the total fluence was, approximately, one photon per initial molecule. 
In experiments 7 - 10 we used $^{13}$CH$_{4}$ instead of CH$_{4}$ to confirm the results found in experiments 3 - 6 and to reject fake positives 
due to contamination. 
As in Sect. \ref{et}, we used the FTIR spectrometer to monitor the ice composition after every irradiation interval and to detect new IR features 
corresponding to the formation of photoproducts. 
The presence of these new species could be subsequently confirmed during the TPD performed at the end of the experiments when necessary. 
Results for the photon-induced chemistry are shown in Sect. \ref{chem_bin}. 
At the same time, the QMS was used during irradiation to detect the molecules desorbing into the gas phase as a consequence 
of the UV photoprocessing. Photon-induced desorption is studied in Sect. \ref{des_bin}. 

\begin{figure*}
\centering
 \includegraphics[width=13.cm]{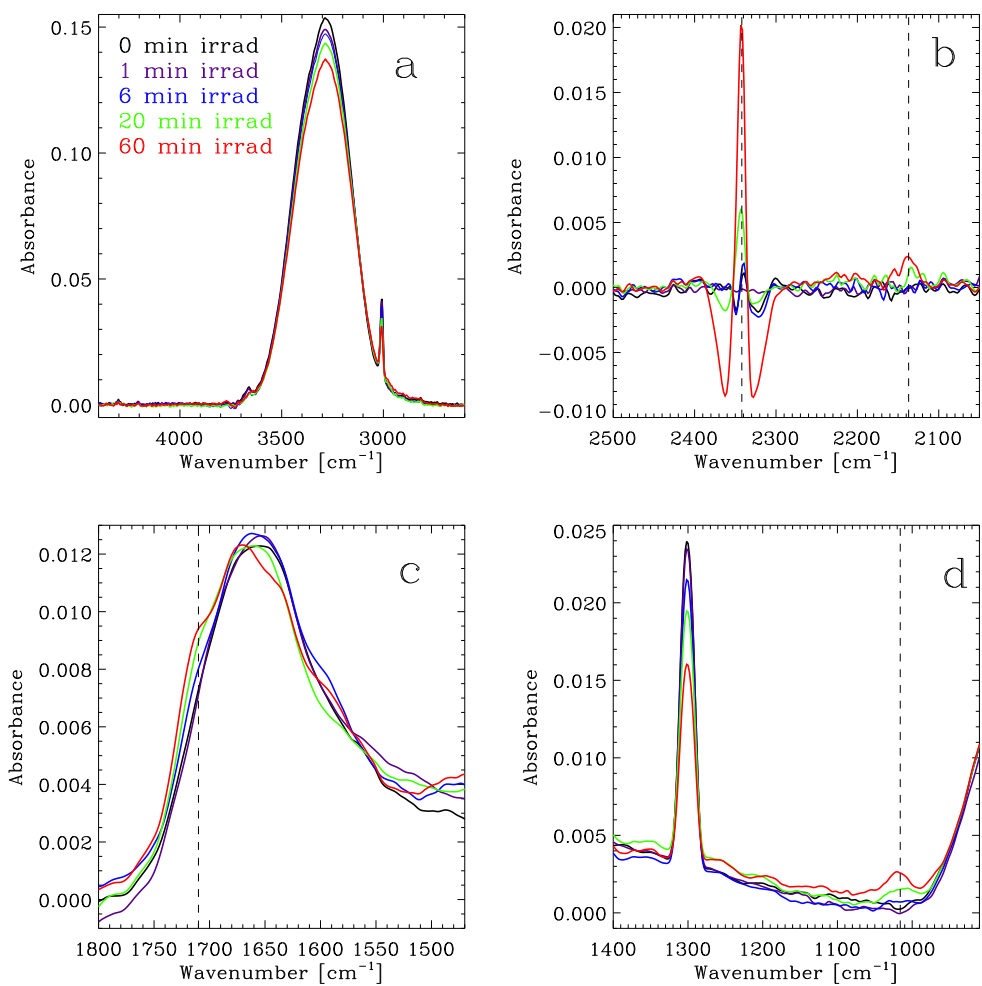}
 \caption{Evolution of the IR spectrum of the H$_{2}$O:CH$_{4}$ ice mixture during UV photoprocessing in experiment 6, see Table \ref{exp_bin}. 
 Results were similar in experiments 3 - 5. 
\textbf{a)} Evolution of the O-H stretching band of H$_{2}$O (broad band peaking at $\sim$3280 cm$^{-1}$),  
and the C-H stretching band of CH$_{4}$ (in the red side of the former). 
\textbf{b)} Formation of CO$_{2}$ and CO 
leads to the appearence of two C-O stretching bands at 2342 cm$^{-1}$ and 2137 cm$^{-1}$, respectively (dashed lines). 
Negative absorbances are due to the atmospheric correction applied by the FTIR software, but the band is due to solid CO$_{2}$, as the peak 
position indicates.
\textbf{c)} Evolution of the O-H deformation band of H$_{2}$O. 
The apearance of a shoulder at $\sim$1710 cm$^{-1}$ is probably due to the formation of H$_{2}$CO. 
\textbf{d)} Evolution of the C-H deformation band of CH$_{4}$ at 1300 cm$^{-1}$. 
The new feature at 1016 cm$^{-1}$ is due to the formation of methanol.
}  
\label{IRbin}
\end{figure*}

\begin{figure*}
\centering
 \includegraphics[width=19.cm]{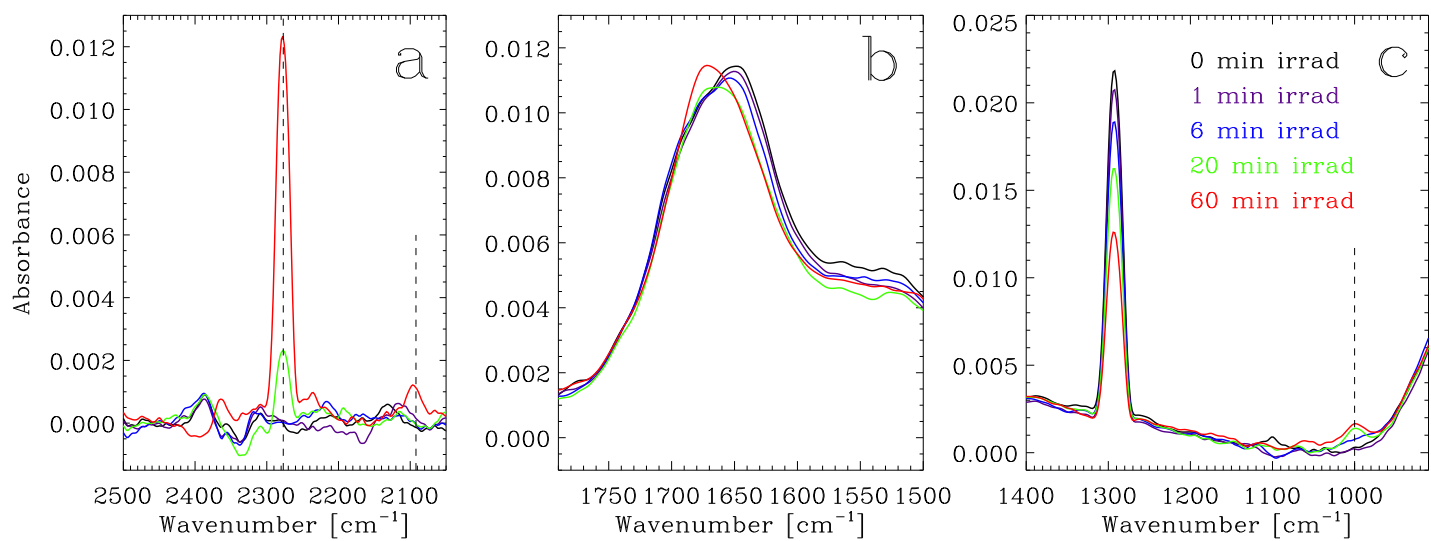}
 \caption{Evolution of the IR spectrum of the H$_{2}$O:CH$_{4}$ ice mixture during UV photoprocessing in experiment 9, see Table \ref{exp_bin}. 
 Results were similar in experiments 7, 8, and 10. 
\textbf{a)} Formation of $^{13}$CO$_{2}$ and $^{13}$CO leads to the 
appearence of two C-O stretching bands at 2277 cm$^{-1}$ and 2092 cm$^{-1}$, respectively (dashed lines). 
\textbf{b)} Evolution of the O-H deformation band of H$_{2}$O. 
\textbf{c)} Evolution of the C-H deformation band of CH$_{4}$ at 1292 cm$^{-1}$. 
The new feature at 1000 cm$^{-1}$ is due to the formation of methanol.
} 
\label{IRbinis}
\end{figure*}

\begin{figure*}
\centering
 \includegraphics[width=13.cm]{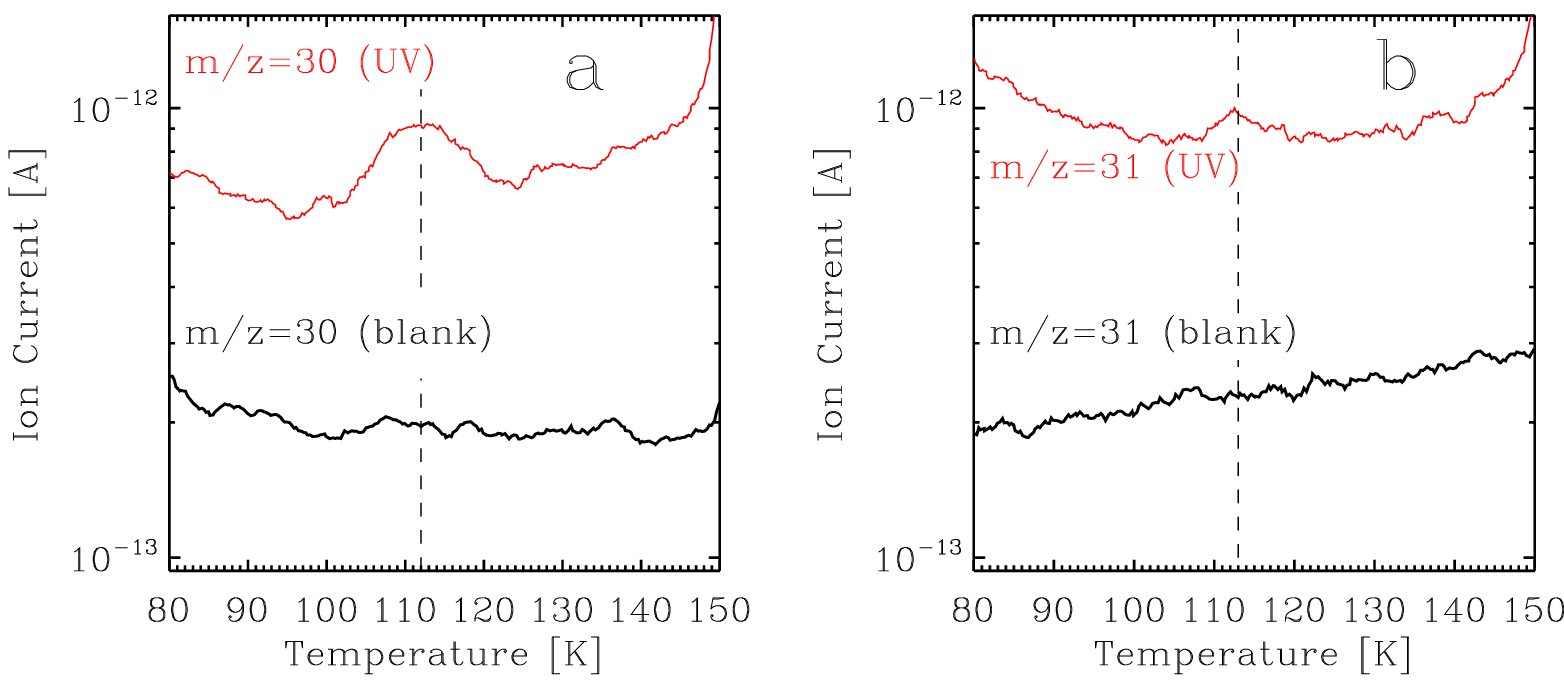}
 \caption{
\textbf{a)} TPD curve of the m/z=30 mass fragment, corresponding to the molecular ion of H$_{2}$CO, 
after irradiation in experiment 3 (red) compared to the blank (black). 
A desorption peak at $\sim$112 K (dashed line) confirms the formation of formaldehyde. 
\textbf{b)} TPD curve of the m/z=31 mass fragment, corresponding to the molecular ion of H$_{2}^{13}$CO, 
after irradiation in experiment 7 (red) compared to the blank (black). 
A desorption peak at $\sim$113 K (dashed line) confirms the formation of formaldehyde. 
} 
\label{TPD_h2co_binaria}
\end{figure*}

\subsubsection{Photon-induced chemistry of a H$_{2}$O:CH$_{4}$ ice analog}
\label{chem_bin}

Figure \ref{IRbin} shows the evolution of the mid-IR spectrum of the ice sample in experiment 6 between 4400 cm$^{-1}$ and 910 cm$^{-1}$, and  
Fig. \ref{IRbinis} shows the mid-IR spectra between 2500 cm$^{-1}$ and 910 cm$^{-1}$ of the ice sample in experiment 9. 
Results were similar for the rest of the experiments. 
As expected, some of the IR features are shifted in the experiments with $^{13}$CH$_{4}$. 
In particular, the C-H deformation band of CH$_{4}$ shifts from 1300 cm$^{-1}$ in Fig. \ref{IRbin}d to 1292 cm$^{-1}$ in Fig. \ref{IRbinis}c. 
Bands of C-bearing photoproducts are also shifted in a similar way (see below).  

The initial components of the ice analogs were monitored with the O-H stretching band at $\sim$3280 cm$^{-1}$ corresponding to water molecules 
(Fig. \ref{IRbin}a),
and in the C-H deformation band mentioned above corresponding to methane (Figures \ref{IRbin}d and \ref{IRbinis}c).  
The intensity of these bands gradually decrease with continuing irradiation, leading to a loss of $\sim$11\% of initial H$_{2}$O and $\sim$35\% 
of initial CH$_{4}$ molecules at the end of experiments 3 - 6 (Table \ref{exp_bin}). 
Loss of water was similar within 10\% in experiments with $^{13}$CH$_{4}$, while loss of methane was about 40\% higher,  
with no obvious explanation. 

In this series of experiments, formation of CH$_{3}$OH during UV photoprocessing of the ice analogs was observed, thanks 
to the detection of the C-O stretching band at 1016 cm$^{-1}$ in Fig. \ref{IRbin}d and at 1000 cm$^{-1}$ in Fig. \ref{IRbinis}c. 
The column density of methanol in the ice at the end of irradiation in experiments 3 - 6 accounts for  $\sim$4\% of the initial CH$_{4}$ column 
density (Table \ref{exp_bin}). 
Formation of methanol took place to a lower extent in experiments 7 - 10.  
It did not reach the sensitivity limit of our FTIR in experiment 7 (see Table \ref{exp_bin}), but it was detected by the QMS during TPD 
(not shown). 
On the other hand, the relatively strong O-H bending band that is due to water molecules
prevented us from clearly detecting the C=O stretching mode of 
H$_{2}$CO in Figs. \ref{IRbin}c and \ref{IRbinis}b. 
A shoulder at $\sim$1715 cm$^{-1}$ is observed on the blue side of the O-H bending band in Fig. \ref{IRbin}c, 
while a shift in the peak frequency of this band is observed in Fig. \ref{IRbinis}b. 
Formation of H$_{2}$CO was confirmed during the TPD performed at the end of the experiments. 
A peak near the desorption temperature of pure H$_{2}$CO \citep[$\sim$112 K, ][]{noble12} was observed for the mass fragments m/z=30, 
corresponding to H$_{2}$CO$^{+}$ in experiments 3 - 6 (Fig. \ref{TPD_h2co_binaria}a), 
and m/z=31, corresponding to H$_{2}^{13}$CO$^{+}$ in experiments 7 - 10 (Fig. \ref{TPD_h2co_binaria}b). 
However, photoproduction of formaldehyde could not be quantified using the IR spectra, since its IR band was barely detected. 

Photochemistry was simpler in this series of experiments.
In addition to CH$_{3}$OH and H$_{2}$CO, only CO and CO$_{2}$ were detected in the IR spectra. 
The C-O stretching mode of these molecules is observed at 2137 cm$^{-1}$ and 2342 cm$^{-1}$, respectively, in Fig. \ref{IRbin}b. 
The C-O stretching bands of $^{13}$CO and $^{13}$CO$_{2}$ are redshifted to 2092 cm$^{-1}$ and 2277 cm$^{-1}$ in Fig. \ref{IRbinis}a, 
respectively, in fair agreement with the values reported in \citet{gerakines95}.

\begin{table*}
\centering
\caption{Values used in Eq. \ref{eqmscal} to convert integrated QMS signals in experiments 7 - 10 (see Table \ref{exp_bin}) into 
column densities of desorbed molecules.}
\label{param2}
\begin{tabular}{ccc}
\hline
\hline
Factor&H$_{2}^{13}$CO&$^{13}$CO\\
\hline
$\sigma^{+} (mol)$ ($\AA^{2}$)$^{a}$&4.140&2.516\\
fragment&H$^{13}$CO$^{+}$&$^{13}$CO$^{+}$\\
$m/z$&30&29\\
$I_{F} (z)$&1$^{b}$&1$^{b}$\\
$F_{F} (m)$&0.549$^{c}$&0.949$^{d}$\\
$k^{*}_{QMS} \cdot S (m/z)$ (A mbar$^{-1}$ $\AA^{-2}$)$^{e}$&9.30 x 10$^{14}$&9.77 x 10$^{14}$\\ 
$k^{*}_{QMS} \cdot S (m/z)$ (A mbar$^{-1}$ $\AA^{-2}$)$^{f}$&1.29 x 10$^{15}$&1.36 x 10$^{15}$\\ 
$k^{*}_{QMS} \cdot S (m/z)$ (A mbar$^{-1}$ $\AA^{-2}$)$^{g}$ &1.77 x 10$^{15}$&1.89 x 10$^{15}$\\
$S_{rel} (mol)^{h}$&0.973&0.991\\
\hline
\end{tabular}
\begin{list}{}
\item $^{a}$ We used the values corresponding to H$_{2}^{12}$CO and $^{12}$CO, extracted from the online NIST database, 
as an approximation.\\ 
\item $^{b}$ A value of 1 was adopted, assuming that no double ionization of the molecules takes place.\\
\item $^{c}$ We used the value corresponding to the fragment H$^{12}$CO$^{+}$, extracted from the online NIST database, 
as an approximation.\\
\item $^{d}$ We used the value corresponding to the molecular ion $^{12}$CO$^{+}$, extracted from the mass spectra library of the QMS 
software, as an approximation.\\
\item $^{e}$ Calculated for our QMS and valid for experiment 7 (see Sect. \ref{lab})\\
\item $^{f}$ For experiment 8 (see Sect. \ref{lab})\\
\item $^{g}$ For experiment 9 and 10 (see Sect. \ref{lab})\\
\item $^{h}$ Calculated with equation \ref{eqmscorr}.\\
\end{list}
\end{table*}

\subsubsection{Photon-induced desorption from a H$_{2}$O:CH$_{4}$ ice analog}
\label{des_bin}

\begin{figure*}
\centering
 \includegraphics[width=18.cm]{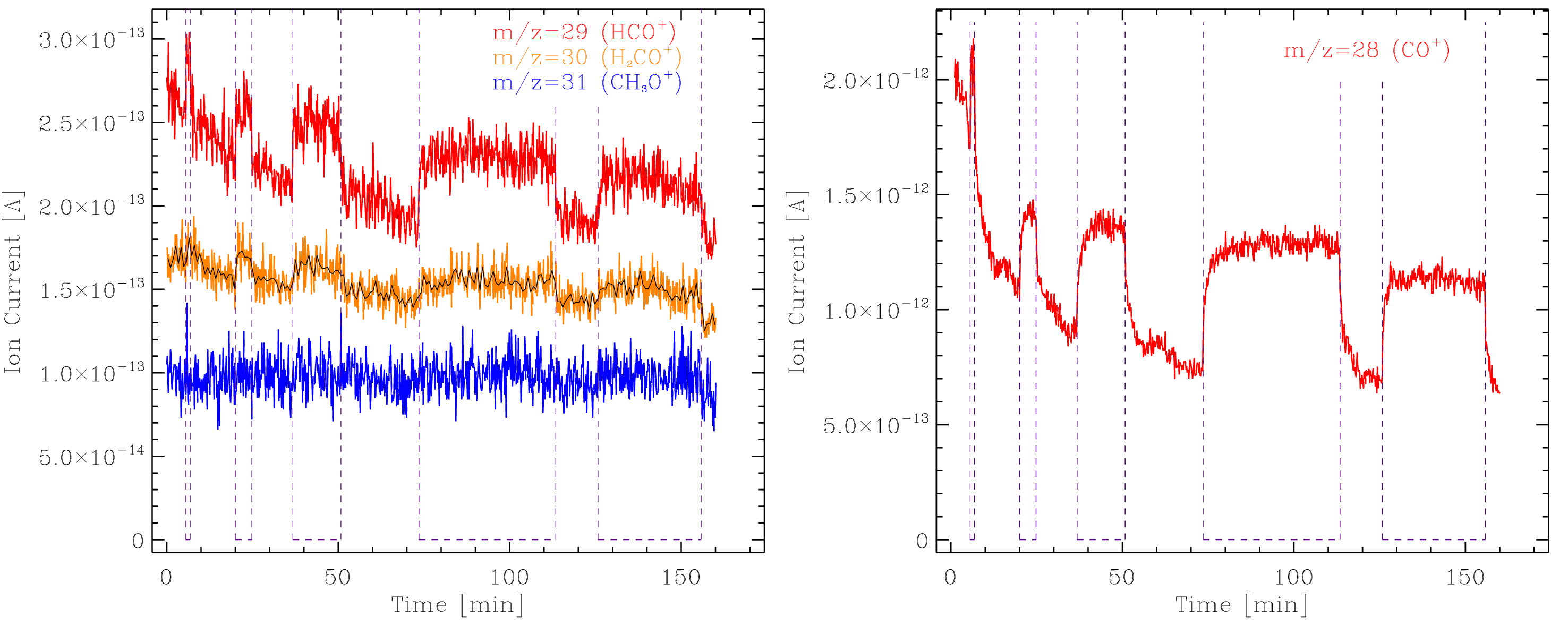}
 \caption{Photon-induced desorption of formaldehyde observed with mass fragments m/z=29 and m/z=30 (\textit{left panel}),  
and carbon monoxide (m/z=28 in \textit{right panel}) 
during UV irradiation of a H$_{2}$O:CH$_{4}$ ice analog in experiment 5. 
Results were similar in exp. 3, 4, and 6. 
Ion currents of different mass fragments are offset for clarity. 
The black solid line represent an average of the corresponding ion current, shown to better evaluate the shape of the signal. 
Irradiation intervals are indicated with dashed vertical lines. 
} 
\label{irradiacion_binaria}
\end{figure*}

\begin{figure*}
 \includegraphics[width=18.cm]{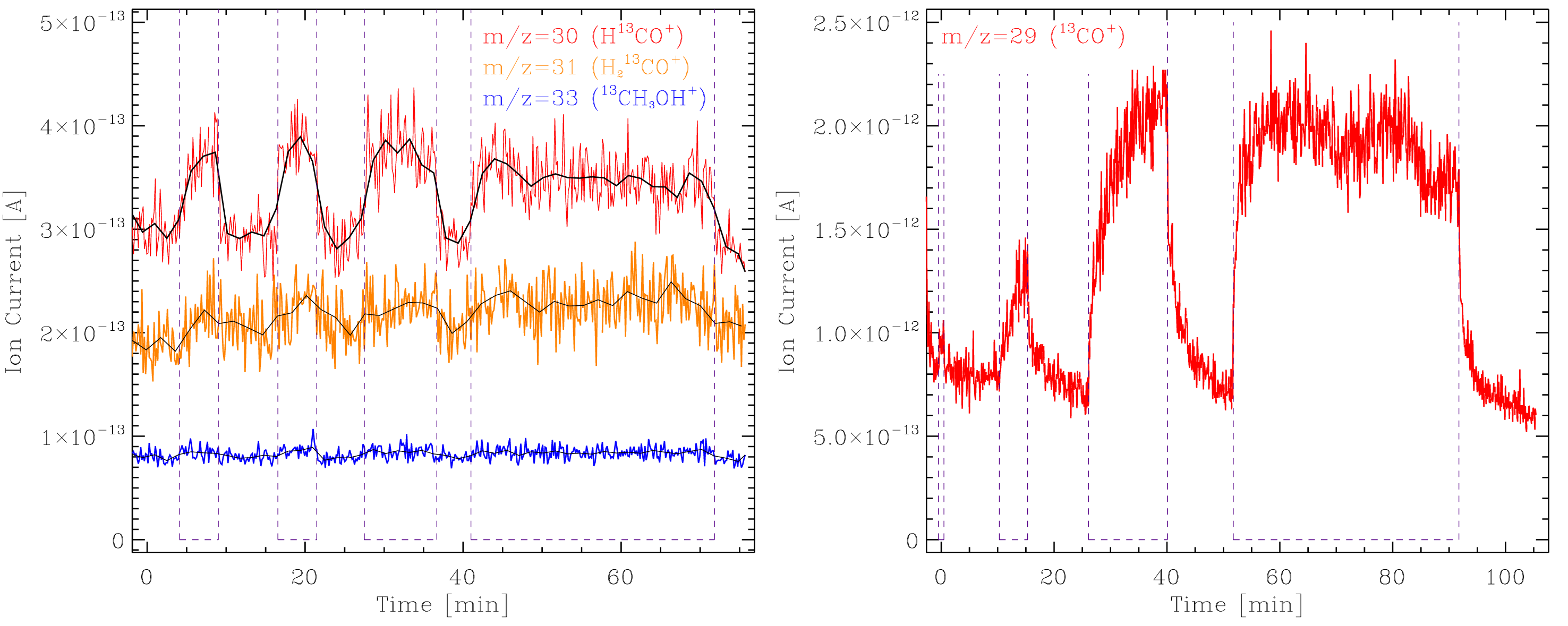}
 \caption{Photon-induced desorption of H$_{2}^{13}$CO (m/z=30 and m/z=31 in \textit{left panel}) and $^{13}$CO (m/z=29 in \textit{right panel}) 
during UV irradiation of a H$_{2}$O:$^{13}$CH$_{4}$ ice analog in experiments 10 and 7, respectively. 
Results were similar for experiments 8 and 9.  
Ion currents of different mass fragments are offset for clarity. 
Black solid lines represent an average of the corresponding ion currents, as shown to better evaluate the shape of the signals. 
Irradiation intervals are indicated with dashed vertical lines. 
} 
\label{irradiacion_binariais}
\end{figure*}

As in Sect. \ref{des_et}, the QMS allowed us to detect desorption of the photoproducts observed in Sect. \ref{chem_bin} during irradiation 
of the ice analogs, monitoring the mass ion fragments selected in Sect. \ref{lab}. 
Results are shown in Figs. \ref{irradiacion_binaria} and \ref{irradiacion_binariais}.
As explained in Sect. \ref{lab}, 
mass fragments in experiments 7 - 10 were 1 amu higher than in exp. 3 - 6 because of the presence of $^{13}$C in the molecules. 
Photon-induced desorption yields were calculated for experiments 7 - 10 using the parameters shown in Table \ref{param2}, 
since experiments with isotopically labeled molecules are more reliable thanks to the lack of contamination effects. 
Compared to the m/z=31 fragment that corresponds to the molecular ion H$_{2}^{13}$CO$^{+}$, 
the H$^{13}$CO$^{+}$ fragment (m/z=30) displayed a more intense signal with a lower noise level. 
Therefore, 
the latter was used 
to quantify the formaldehyde photon-induced desorption yield (the molecular ion was used in exp. 1 - 2). 
Proportionality constant $k_{CO}$ varied between 1.35 x 10$^{-10}$ A min ML$^{-1}$ and 1.90 x 10$^{-10}$ A min ML$^{-1}$ for experiments 7 - 10. 
Table \ref{ypdbin} shows the photon-induced desorption yields measured in experiment 9,  averaged for every irradiation interval. 
Results were similar in experiments 7, 8, and 10.  

No clear photon-induced desorption was observed for the mass fragments m/z=31 (left panel of Fig. \ref{irradiacion_binaria}) 
and m/z=33 (left panel of Fig. \ref{irradiacion_binariais}) corresponding to methanol molecules in experiments 3 - 6 and 7 - 10, respectively. 
On the other hand, a constant rise was detected during irradiation in experiments 3 - 6 for mass fragments m/z=29 and m/z=30   
(left panel of Fig. \ref{irradiacion_binaria}), suggesting photochemidesorption of formaldehyde. 
The same behavior is observed in mass fragments m/z=30 and m/z=31 in experiments 7 - 10 (left panel of Fig. \ref{irradiacion_binariais}).  
This confirms photochemidesorption of formaldehyde, since the fragments of H$_{2}^{13}$CO are 1 amu higher than those of H$_{2}$CO.  
Averaged photodesorption yields of H$_{2}^{13}$CO in Table \ref{ypdbin} are 
slightly lower than the upper limits derived 
in Sect. \ref{des_et} for the pure C$_{2}$H$_{5}$OH ice photoprocessing experiments. 

Desorption of CO molecules during irradiation is detected in the righthand panel of Fig. \ref{irradiacion_binaria} with the mass fragment m/z=28. 
The DIET-like behavior of the photon-induced desorption is more clearly observed with the mass fragment m/z=29 in the righthand panel of 
Fig. \ref{irradiacion_binariais}. 
The increasing photodesorption yield of $^{13}$CO in Table \ref{ypdbin} is one order of magnitude lower than that found in Sect. \ref{des_et}. 
This is probably due to the much lower formation rate found for the CO molecules in this series of experiments. 
After a fluence of 7.2 photons cm$^{-2}$, only $\sim$0.5\% of the initial column density had formed CO in experiments 3 - 10, 
while the value in experiments 1 and 2 is $\sim$13\% 
(see the low intensity of the C-O stretching band corresponding to CO in Figs. \ref{IRbin}b and \ref{IRbinis}a compared to Fig. \ref{IRetanol}b). 
Photon-induced desorption of CO$_{2}$ was not observed in experiments 3 - 10 (not shown in Figures \ref{irradiacion_binaria} and 
\ref{irradiacion_binariais}). 

\begin{table*}
\centering
\caption{Evolution of the photodesorption yields during UV irradiation of a binary H$_{2}$O:$^{13}$CH$_{4}$ ice analog in exp. 9, 
see Table \ref{exp_bin}. 
Results were similar in experiments 7, 8, and 10.}
\label{ypdbin}
\begin{tabular}{cccc}
\hline
\hline
Irradiation period&Fluence$^{a}$&$Y_{pd}$ (H$_{2}^{13}$CO)$^{b}$&$Y_{pd}$ ($^{13}$CO)$^{b}$\\
(min)&(photons cm$^{-2}$)&10$^{-5}$ (molecules/incident photon)
&10$^{-4}$ (molecules/incident photon)\\
\hline
0 - 1&1.2 x 10$^{16}$&4.3&0.8\\
1 - 6&7.2 x 10$^{16}$&3.9&0.9\\
6 - 20&2.4 x 10$^{17}$&4.8&1.5\\
20 - 60&7.2 x 10$^{17}$&4.4&4.9\\
\hline
\end{tabular}
\begin{list}{}
\item $^{a}$ Total fluence at the end of the irradiation period. 
\item $^{b}$ Averaged for each irradiation period in experiment 9. 
Photon-induced desorption yield values could be different by a factor 2 due to the uncertainties in all the parameters 
of equation \ref{eqmscal}.
\end{list}
\end{table*}

\section{Astrophysical implications}
\label{imp}
In cold dense interstellar and circumstellar regions, molecules form ice mantles upon freeze-out onto cold dust grains. 
These ice mantles are mainly observed in the near- to far-IR region of the spectrum \citep[and ref. therein]{boogert15}.  
Solid H$_{2}$O is the most abundant species observed in interstellar and circumstellar ices with an average abundance of 
$\sim$ 4 $\times$ 10$^{-5}$ relative to N$_{H}$ \citep{boogert15}. 
Several other species have been confirmed to be present in ice mantles, in particular, CO, CO$_{2}$, CH$_{3}$OH, NH$_{3}$ and CH$_{4}$. 
Solid methane has been detected with abundances ranging from 0.4\% to 11\% relative to water ice \citep[and ref. therein]{boogert15}. 
In the second series of experiments, we used binary H$_{2}$O:CH$_{4}$ ice analogs with a methane abundance that is somehow higher ($\sim$33\%). 
More species have been tentatively identified in interstellar and circumstellar ices, some of them observed in comets. 
As mentioned in Sect. \ref{intro}, C$_{2}$H$_{5}$OH has been suggested as one of the possible carriers of the 7.24 $\mu$m IR band 
\citep{schutte99}. 
\citet{schriver07} derived an upper limit to the abundance of ethanol in interstellar ices of 1.2\% relative to water ice from ISO observations. 
In addition, the presence of solid ethanol is needed to explain gas-phase abundances in dense star-forming regions 
that cannot be reproduced by pure gas-phase chemical models.  
Pure ethanol ice samples were used for our first series of experiments. 

Ice mantles can be energetically processed by cosmic rays, UV, and X-ray photons, and also thermally in regions around protostars. 
Energetic processing of ices leads to several effects, including photon-induced chemistry and photon-induced desorption as studied in this work. 
Solid-phase chemistry has been widely proposed as a possible source of molecules in the ISM. 
In some cases, formation in ice mantles followed by nonthermal desorption needs to be invoked to explain the observed abundances of some 
species in dense cores and low UV-field illuminated PDRs. This is the case of the targeted molecules of this work, CH$_{3}$OH and H$_{2}$CO 
(see Sect. \ref{intro}). 
Photoprocessing in particular takes place in dense cores thanks to the secondary UV field generated by the interaction of H$_{2}$ molecules 
in the gas phase with cosmic rays, leading to a photon flux of $\sim$10$^{4}$ photons cm$^{-2}$ s$^{-1}$ \citep{pestellini92,shen04}. 
On the other hand, PDRs are illuminated by nearby massive stars with a UV flux of $\sim$10$^{11}$ eV cm$^{-2}$ s$^{-1}$ for a 
low UV-field illuminated PDR as for the Horsehead PDR \citep{pety12}.
In our experiments, we used a MDHL with a photon flux of $\sim$2 $\times$ 10$^{14}$ photons cm$^{-2}$ s$^{-1}$ and an 
average photon energy of 8.6 eV (see Sect. \ref{lab}). 
Therefore, the ice samples in our experimental simulations experience the same total fluence as the ice mantles during the typical lifetime 
of dense cores ($\sim$3 $\times$ 10$^{17}$ photons cm$^{-2}$ for a cloud lifetime of 10$^{6}$ years) after 30 minutes of irradiation. 
At the same time, the UV flux experienced by the ice samples in the experiments is about four orders of magnitude higher than that present 
in low-illuminated PDRs. 

As mentioned in Sect. \ref{intro}, gas-phase abundances of CH$_{3}$OH in dense cores and PDRs where thermal desorption is inhibited cannot be 
reproduced by gas-phase chemical models, and solid chemistry followed by nonthermal desorption needs to be invoked. 
Methanol formation was not observed during irradiation of a pure ethanol ice sample, but it was formed in the H$_{2}$O:CH$_{4}$ ice 
irradiation 
experiments, accounting for $\sim$1\% of the total initial column density. However, photon-induced desorption of methanol was not observed 
in this second series of experiments, and therefore its abundance in these regions remains an open question in astrochemistry. 

On the other hand, formaldehyde was formed in both series of experiments. 
Photochemidesorption of this species was detected with a yield of 
$\le$6 x 10$^{-5}$ (molecules/incident photon) in the case of the pure ethanol 
ice experiments, and $\sim$4.4 x 10$^{-5}$ (molecules/incident photon) when the H$_{2}$O:CH$_{4}$ ice analogs were 
photoprocessed. 
The latter value (more accurate, and closer to the astrophysical scenario) could be used in gas-grain chemical models, taking
photon-induced desorption from dust grains and subsequent gas-phase reactions into account, in order to test whether photochemidesorption of 
formaldehyde as observed in this work is able to account for all the formaldehyde detected in low UV-field illuminated PDRs 
or if an additional source of formaldehyde is needed. 

\section{Conclusions}
\label{conclusiones}

We have explored two UV-induced formation pathways and subsequent desorption for CH$_{3}$OH and H$_{2}$CO, 
which are two simple organic molecules that are ubiquitously detected in the ISM. 
Photoproduction of these species was searched for with an IR spectrometer during UV irradiation of the ice samples. 
Photon-induced desorption of several photoproducts was observed directly in the gas phase by means of mass spectrometry. 
Calibration of our QMS allowed us to quantify the photon-induced desorbing molecules and calculate desorption yields. 

In a first series of experiments, a top-down approach to the formation of methanol and formaldehyde was used 
when studying the photodissociation of a pure ethanol ice. 
Seven photoproducts were detected: H$_{2}$O, CO, CO$_{2}$, CH$_{4}$, C$_{2}$H$_{6}$, H$_{2}$CO, and CH$_{3}$CHO. 
Photon-induced desorption of four of the photoproducts was observed during photoprocessing of the ice sample.  
The increasing desorption yield observed for H$_{2}$O, CO, and CH$_{4}$ suggests that the DIET mechanism was active for these species. 
In particular, the photon-induced desorption yield of water and methane reached constant values of 
$\le$2.6 $\times$ 10$^{-4}$ molecules/incident photon and $\sim$3.3 $\times$ 10$^{-4}$ molecules/incident photon, 
respectively, after a fluence of 1.8 $\times$ 10$^{17}$ photons cm$^{-2}$, while that of carbon monoxide continued increasing, reaching a value 
of 1.5 $\times$ 10$^{-3}$ molecules/incident photon after a total fluence of 7.2 $\times$ 10$^{17}$ photons cm$^{-2}$. 
On the other hand, formaldehyde and/or ethane molecules were observed to photochemidesorb with a constant desorption yield of 
$\sim$6 $\times$ 10$^{-5}$ molecules/incident photon in case the QMS m/z=29 signal was only due to H$_{2}$CO 
molecules. 

In a second series of experiments, a bottom-up approach was explored, using a more realistic water-reach ice analog containing methane. 
Photoprocessing of H$_{2}$O:CH$_{4}$ ice samples with a 3:1 ratio led to a simpler photochemical network , 
and only four photoproducts were detected: CO, CO$_{2}$, CH$_{3}$OH, and H$_{2}$CO. 
Only CO and H$_{2}$CO were observed to desorb upon UV irradiation of the ice mixtures. 
The photon-induced desorption yield of carbon monoxide was found to increase during photoprocessing up to 
4.9 $\times$ 10$^{-4}$ molecules/incident photon 
after a total fluence of 7.2 $\times$ 10$^{17}$ photons cm$^{-2}$, suggesting that the desorption was driven by the DIET mechanism. 
Photochemidesorption of formaldehyde was also detected, with a constant yield of $\sim$4.4 x 10$^{-5}$ molecules/incident photon. 
These results were confirmed using H$_{2}$O:$^{13}$CH$_{4}$ ice analogs. 

Methanol was only formed during the second series of experiments, but no significant photon-induced desorption was observed. 
So far, there has been no experimental evidence of an efficient nonthermal desorption mechanism that explained the gas-phase methanol abundances 
in dense cores 
and low UV-field illuminated PDRs, where thermal desorption from ice mantles is inhibited, and gas-phase chemical models fail to account for 
all the detected methanol.  
On the other hand, formaldehyde was photoproduced in both series of experiments. Photochemidesorption took place with a 
similar yield in both cases. 
While gas-phase chemical models are able to reproduce the formaldehyde abundances in dense cores, formation in the solid phase and 
subsequent desorption to the gas phase is needed in the case of low UV-field illuminated PDRs. 
The photon-induced desorption yield found in this work can be used in gas-grain chemical models to test that the observed column densities of 
formaldehyde in low UV-field illuminated PDRs can be explained by the proposed photon-induced formation and subsequent desorption process. 

\begin{acknowledgements}
We are grateful to Javier Manzano-Santamar\'ia for his support on the experiments. 
Special thanks go to Marcelino Ag\'undez for useful discussions. 
This research was financed by the Spanish MINECO under project AYA2011-29375. R.M.D. benefited from a FPI grant from Spanish MINECO. 
\end{acknowledgements}

\end{document}